\documentclass[lettersize,journal]{IEEEtran}
\usepackage{amsmath}
\usepackage{hyperref}
\usepackage{paralist}
\usepackage{algorithm}
\usepackage{algorithmicx}
\usepackage{algpseudocode}
\newtheorem{theorem}{Theorem}
\newtheorem{lemma}{Lemma}
\newtheorem{assumption}{Assumption}

\usepackage{subcaption}
\usepackage{graphicx}

\begin{document}

\title{orb-QFL: Orbital Quantum Federated Learning}

\author{Dev Gurung and Shiva Raj Pokhrel, Senior Member, IEEE 
\thanks{Authors are with the QUANTIMA Research, School of IT, Deakin University, Geelong, Australia}
\thanks{Manuscript received April 19, 2025; revised August 16, 2026.}}
\markboth{Journal of \LaTeX\ Class Files,~Vol.~14, No.~8, August~2026}%
{Shell \MakeLowercase{\textit{et al.}}: A Sample Article Using IEEEtran.cls for IEEE Journals}


\maketitle



\begin{abstract}
Recent breakthroughs in quantum computing present transformative opportunities for advancing Federated Learning (FL), particularly in non-terrestrial environments characterized by stringent communication and coordination constraints. In this study, we propose orbital QFL, termed \textit{orb-QFL}, a novel quantum-assisted Federated Learning framework tailored for Low Earth Orbit (LEO) satellite constellations. Distinct from conventional FL paradigms, termed orb-QFL operates without centralized servers or global aggregation mechanisms (e.g., FedAvg), instead leveraging quantum entanglement and local quantum processing to facilitate decentralized, inter-satellite collaboration. This design inherently addresses the challenges of orbital dynamics, such as intermittent connectivity, high propagation delays, and coverage variability. The framework enables continuous model refinement through direct quantum-based synchronization between neighboring satellites, thereby enhancing resilience and preserving data locality. To validate our approach, we integrate the Qiskit quantum machine learning toolkit with Poliastro-based orbital simulations and conduct experiments using Statlog dataset.
\end{abstract}

\maketitle


\section{Introduction}
Today, there is an increase in data-driven applications. 
Along with it, due to the need for better wireless communication, sixth-generation (6G) mobile systems enhanced by AI have substantial research interest.
One of the best candidates considered for the 6G network is low-Earth orbit (LEO) satellites \cite{chenSatelliteBasedComputingNetworks2022}. Numerous LEO satellite constellations are deployed for Earth observation and satellite-ground communication, gathering vast amounts of imagery and sensor data \cite{yangCommunicationEfficientSatelliteGroundFederated2024}.
 With these data, smart satellite applications can be enhanced to navigate and mitigate various global challenges, such as real-time disasters.

Satellite communication networks are also considered an effective approach to addressing connectivity problems in many remote areas~\cite{pokhrel2024data}.
The influx of IoT devices is expected to reach 30 billion by 2030. However, the reach is limited due to various challenging problems.
To achieve global coverage and seamless connectivity to enable AI tasks,
satellite communications (SATCOM) networks are considered a practical complementary alternative to terrestrial networks \cite{chenSatelliteBasedComputingNetworks2022}.
\begin{figure}[!htbp]
\centering
\includegraphics[width=0.6\columnwidth]{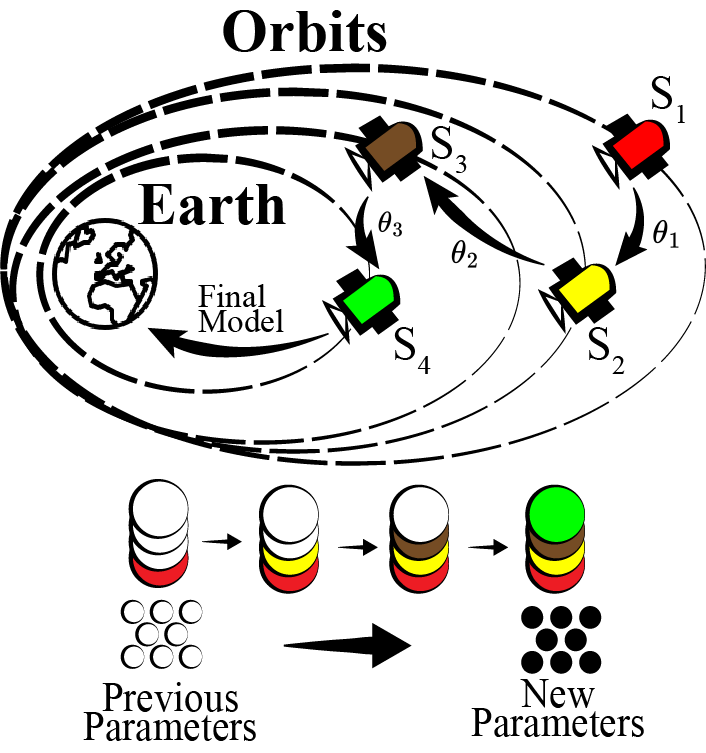}
\caption{\textit{Overview of orb-QFL framework for LEO satellite constellation demonstrating proposed learning approach.}}
\label{fig:overview}
\end{figure}
There has been increasing interest in research on operationalizing federated learning (FL) with SATCOM to facilitate collaborative training of machine learning models among numerous LEO satellites \cite{elmahallawyCommunicationEfficientFederatedLearning2024}.
However, implementation of FL in the SATCOM environment faces various challenges, such as delay in FL training which can extend to days or weeks \cite{elmahallawyCommunicationEfficientFederatedLearning2024}.
Other challenges include the requirement for a clear line of sight between the antenna and the satellite due to zero blind spots in the satellites, constellation challenges, design challenges, high cost etc.
Similarly, depending upon the type of satellites such GEO, MEO or LEO, there can be subtantial propagation delays unsuitable for time-critical communications for GEO satellites, visibility requirement of at least four satellites for MEO satellites and demand for large number of LEO satellites in order to cover significantly smaller regions.

Quantum federated learning (QFL) is the field of study which focuses on providing quantum advantage to federated learning (FL) in quantum computing setting \cite{ren_towards_2023, gurungPerformanceAnalysisDesign2025}.
With numerous works in the field of QFL \cite{gurungChainedContinuousQuantum2025, gurungBQFLMetaverse, chen_introduction_2024} and
its promising computational capacities and improved optimization, it is poised perfectly in distributed quantum machine networks such as SATCOM.

However, implementing a QFL framework in satellite constellations presents significant challenges both quantum challenges and unique constraints of SATCOM. 
Many of the latest studies aimed at implementing FL across SATCOM for collaborative machine learning training emphasize and perform the study on a server-centric approach with classical computing capabilities. 
Unlike these studies, our research departs considerably by focusing on integrating quantum advantage to SATCOM through QFL. 

\textbf{Contribution.}
Our main contributions are outlined below.
\begin{enumerate}
    \item We pioneer a decentralized Quantum Federated Learning paradigm for LEO satellite constellations that eliminates centralized infrastructure. Our framework facilitates direct quantum-empowered model refinement between satellites, completely bypassing terrestrial servers and global aggregation mechanisms (e.g., FedAvg).
    
    \item We establish rigorous theoretical guarantees and demonstrate operational superiority via extensive experiments incorporating orbital dynamics, intermittent visibility, and transmission constraints. Results confirm significant performance gains in convergence speed, communication efficiency, and resilience compared to classical federated learning approaches.
\end{enumerate}


\section{Problem Formulation}
We consider a SATCOM system that consists of LEO constellation without any server communication. 
Somehow, similar to a peer-to-peer approach, our problem formulation is based on a continuous training approach without two-way communication between server and clients as in the traditional FL approach \cite{gurungChainedContinuousQuantum2025}.
Although the design can be varied in various ways, we believe that this problem formulation can play a critical role in new directions toward different FL frameworks for the application scenarios.
Suppose that there are $k$ LEO satellites $\{s_1, s_2,... , s_k \}$ deployed  orbiting equidistant from each other. 
Each node has its own local dataset $\delta_s = \{(x_i, y_i)\}^{\delta_s}_{i=1}$ with $\delta_k$ samples.
Then, the goal of federated learning is to collaboratively learn a global model $\theta_g$.
In traditional FedAvg, the global model $\theta_g$ can be obtained by solving the following optimization problem, 

\begin{equation}
    \min_\theta F(\theta) = \sum_{k \in K} q_s F_s(\theta_s)
\end{equation}
where, $q_s = \delta_s/\delta$ is the weight factor whereas, $\delta$ denotes training data and $F_s$ represents the local loss which can be represented as, 
\begin{equation}
    F_s(\theta_s) = \frac{1}{\delta_s} \sum_{(x_i, y_i) \in \delta_s} f_s(\theta_i; x_i, y_i)
\end{equation}
After, this the model $\theta_s$ is sent to the server for aggregation purposes.
After the  aggregation mechanism, the server then broadcasts the global model to the local devices where they update their local models.
\begin{figure}
    \centering
    \includegraphics[width=\columnwidth]{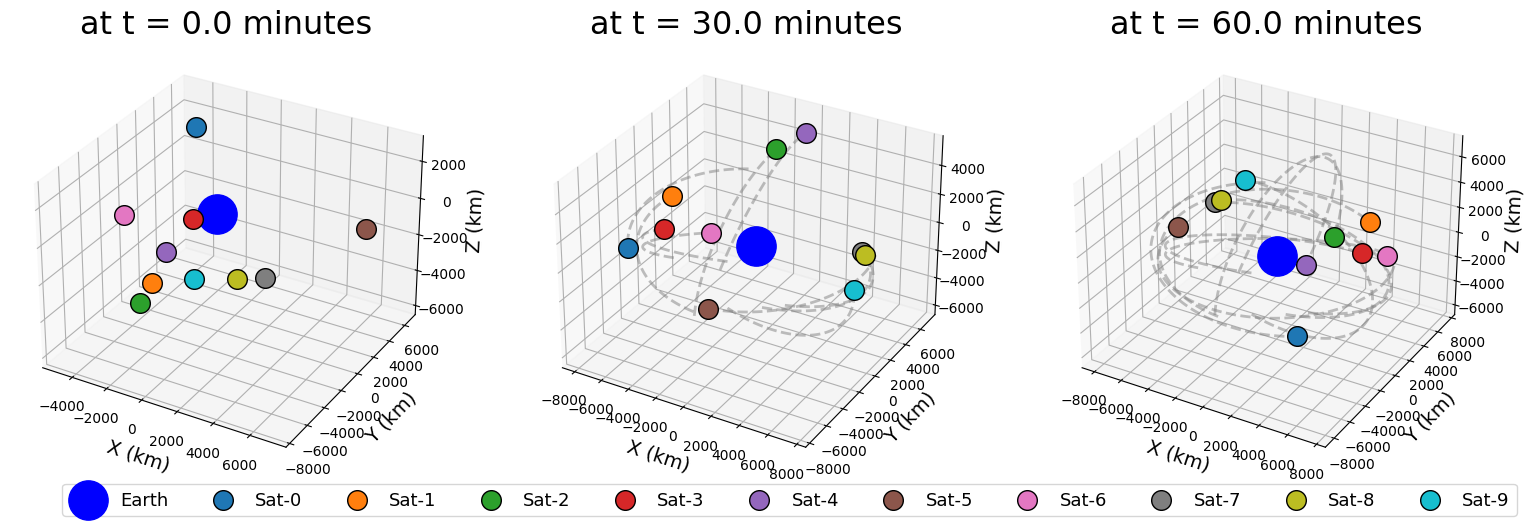}
    \caption{Example showcasing satellite positions and orbits at different times using Poliastro Library.}
    \label{fig:satellite_positions}
\end{figure}
This is where our approach differs. Rather than transmitting the model back to the terrestrial receiver or another server to create a global model $\theta_{g}$, we omit the entire process, which includes \begin{inparaenum}[(i)] \item transmitting the local model $\theta_s$ from the device to the server, \item aggregating models at the server, \item the server broadcasting the updated global model. \end{inparaenum}

Considering various other factors such as intermittent connectivity, limited bandwidth, high latency, and privacy and security concerns, standard FL approaches that use FedAvg or other existing approaches may not be sufficient.  
In standard FL standards, communication between satellites would already be affected by vulnerable communication bottlenecks. 
In such scenarios, it might be more lucrative to apply continuous approaches.
Thus, we aim to improve $\theta_s$ as it is collaboratively improved by all the devices in an orbital fashion. 
The first satellite node $s_1$ sends its updated model after local training to another nearby satellite node, say $s_2$ where the model is further trained.
Thus, we want to optimize a problem which can be stated as, 
\begin{align}
 \overset{\min}{\theta_s^0} 
 & \rightarrow F_s^1(\theta_s^0;x_i, y_i) \notag \\
 & \overset{\theta_x^1}{\rightarrow} F_s^2(\theta_s^1;x_i, y_i) \notag \\
 & \dotsc \notag \\
& \overset{\theta_x^2}{\rightarrow}  F_s^n(\theta_s^{(n-1)};x_i, y_i) \notag \\
& \rightarrow \theta_{g}
\end{align}
until $T$ communication rounds.
At the end of all communication rounds, we aim to achieve the optimal global model $\theta_g$.

\section{Background}
\subsection{QFL}
QFL combines quantum machine learning
with federated learning to take advantage of both paradigms \cite{chen_introduction_2024}. 
In a typical QFL setup, multiple clients \(c_i\) with quantum processors collaboratively train a global
quantum model \(Q(\theta)\) while keeping their local data \(\delta_i\) private. 
Each client \(c_i\) performs local updates on its quantum model parameters \(\theta_i\) using a local loss 
function \(f_i(\theta_i)\), 
and then shares the updated parameters \(\theta_i\) with a central server. 
The server aggregates these parameters to update the global model \(\theta\) using an aggregation rule such as \(\theta = \sum_{i=1}^N w_i \theta_i\), where \(w_i\) represents the weighting factor for the client \(c_i\). This iterative process can be formulated as
$\theta_i^{t+1} = \theta_i^{t} - \eta \nabla_{\theta_i} f_i(\theta_i^{t})$
and 
$\theta^{t+1} = \sum_{i=1}^n w_i \theta_i^{t+1}$
where \(t\) denotes the iteration step, \(\eta\) is the learning rate, and \(\nabla_{\theta_i} L_i(\theta_i^{t})\) is the loss function gradient with respect to local parameters.

\subsection{Satellite Communications}
Satellite communications involve transmitting and receiving radio frequency (RF) 
signals to and from satellites orbiting the Earth \cite{wuClientSelectionSatellite2024}.
It facilitates communication between distant locations on Earth. The high-frequency radio waves used for these connections travel in straight lines and are therefore obstructed by the Earth's curvature. 
Communication satellites are designed to relay and amplify these signals
using a transponder, allowing communication between geographically distant points. 
Communication can be one-way like for television broadcasts, or two-way to handle internet traffic, radio communications navigation signals, etc. 
Bidirectional satellite
communication and cross-links follow a similar process in which two satellites communicate with each other.

\subsection{Satellite Orbits}
Communications satellites usually function within one of three main orbit categories: Geostationary Earth Orbit, Medium Earth Orbit, and Low Earth Orbit.
Satellites in GEO maintain a circular orbit approximately $35,785$ km above the Earth's surface. 
Due to their high altitude and orbiting at Earth's rotational speed, 
they appear stationary to observers on the ground. 
This characteristic enables nearly complete global coverage with only three satellites when they are positioned $120$ degrees apart in longitude. 
Common uses of GEO satellites include weather forecasting, satellite television, 
satellite radio, and other global communication services.
MEO satellites orbit at altitudes ranging from $2,000$ to $36,000$ km above Earth. 
These orbits are typically used for navigation systems such as the global positioning system. 
A constellation of about $32$ MEO satellites is required to ensure continuous global coverage
at any time during the day. 
LEO satellites operate at altitudes ranging from $160$ to $2,000$ km above Earth's surface. 
Recent technological advancements have resulted in the deployment of numerous LEO satellite constellations, 
which are preferred because of their shorter propagation delays, making them ideal for time-sensitive applications. 

\section{Related Work}

In terms of frameworks, Chen et al. \cite{chenSatelliteBasedComputingNetworks2022}
proposed FL in LEO-based satellite communication networks.
Their framework supports massively interconnected devices with the capability of intelligent adaptive learning to reduce expensive traffic in SATCOM.
Elmahallawy et al. \cite{elmahallawyAsyncFLEOAsynchronousFederated2022} proposed an asynchronous FL framework for LEO constellations called AsyncFLEO so that FL efficiency can be improved.
It addresses bottlenecks, i.e. idle waiting in synchronous FL and issues of straggler satellites.
Whereas in another work, Elmahallawy et al. \cite{elmahallawyCommunicationEfficientFederatedLearning2024}
proposed a novel FL-SatCom approach termed NomaFedHAP that uses high-altitude platforms (HAP) as distributed parameter servers so that the visibility of satellites can be further improved.
Pokhrel et al. \cite{pokhrelBlockchainBringsTrust2021a} developed a Blockchain-based framework for continual knowledge sharing and collaborative learning.
The scope of applicability of the proposed approach extends to Unmanned Aerial Vehicles (UAVs).
Yang et al. \cite{yangCommunicationEfficientSatelliteGroundFederated2024} on the other hand proposed an efficient satellite-ground FL framework, Satellite FL to mainly address three challenges i.e. ensuring completion of per-round training within each connection window, maximizing model utility within connection window.


For optimization, Yan et al. \cite{yanConvergenceTimeOptimization2024a} investigated how to minimize the convergence time of decentralized FL with satellites.
They do this through the definition of the number of orbits and the number of satellites for those orbits.
Zhou et al. \cite{zhouDecompositionMetaDRLBased2024} studied the synergy of LEO and FL for future integrated 6G satellite systems.
The authors proposed a decomposition and meta-deep reinforcement learning-based multi-objective optimization algorithm for FL so that it can be adaptive to dynamic satellite-terrestrial environments.

While the majority of research deals with classical federated learning in LEO satellites, our study aims to incorporate quantum machine learning. Furthermore, typical federated learning models rely on a central server structure with some kind of aggregation process. In contrast, our proposed QFL framework eliminates the need for a central server and aggregation mechanisms.


\begin{figure*}[!htb]
    \centering
    \includegraphics[width=0.8\textwidth]{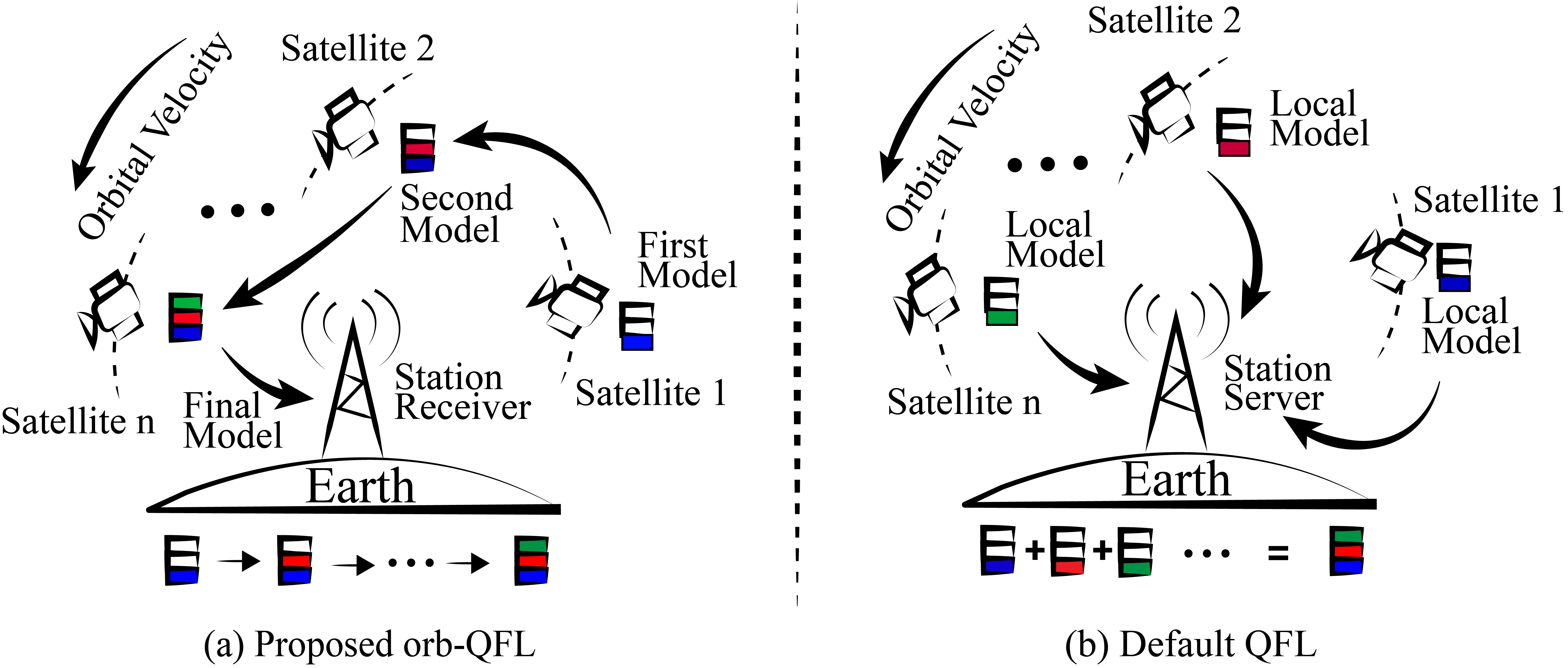}
   \caption{\textit{QFL frameworks: (a) On the left, orb-QFL for LEO Satellite network where satellites communicate with each other to improve distributed model training without a centralized server approach; (b) On the right, the QFL framework with the server has an aggregation mechanism; QFL involves constant two-way communication between server station and satellite along with aggregation mechanism; However, with orb-QFL, the communication just happens between to satellites with probable only once a while communication back to the station especially after completion of all communication rounds.}}
    \label{fig:framework_satQFL}
\end{figure*}

\section{orb-QFL for LEO Satellites}
In this section, we present our proposed orb-QFL framework for the LEO satellite network.
Figure \ref{fig:framework_satQFL} shows an overview of the proposed framework along with the standard QFL approach and how they are different from each other. 
The Algorithm \ref{alg:orb-QFL} shows the basic steps involved in the framework.
The first step in our framework is to identify the peers in the constellation, especially the next one.
Each satellite node $s_i$ in the network consists of its local data set $\delta_i$ that is different from another satellite node $s_{i+1}$ in the network.
To start with, first, we encode the data sets $x$ into their quantum state $|\psi(x)\rangle$ so that they can be used in our quantum machine learning training with each local device.
The training process starts with the first satellite node $s_i$, which initializes its model parameters ${\theta_{satNode}^{VQC}}$ at random.

Once trained on its local model through $VQC_i.fit(\theta_i, |\psi(x)\rangle,y)$, the trained model $\theta_s 
$ 
is transmitted to the other satellite node which is nearby.
Before that, it computes various metrics to maintain line of sight, timing, etc. like computing distance.
The main idea behind the proposed framework is that the next satellite node $x_{i+1}$ after receiving the locally trained model from another nearby satellite node $s_i$ starts training using the new received model $\theta_{i-1}$.
In this way, the node does not have to start training from scratch.

{\small
\begin{algorithm}[!htb]
  \caption{orb-QFL for LEO Satellite Network}
  \label{alg:orb-QFL}
  \centering
  \begin{algorithmic}[1]
  \State \textbf{Initialize:} 
      Satellite id $idx$, features $x$, labels $y$, $Acc_{train}$, $Acc_{test}$, Sampler $Sampler$, Local Iterations $t$, Optimizer $Opt$, Objective Values $obj$, Feature Map $\psi(x)$, Ansatz $\phi$, VQC $VQC$, $n$ number of LEO satellites, local models $\{\theta_i, .., \theta_n$\}, Local datasets  \{$\delta_1, \delta_2,...\delta_n$\}, Communication Rounds $R$, Satellites List $satList[]$, Variational Quantum Classifier $VQC$, File Size $fileS$.
    \Procedure{Data Encoding}{\textit{x}}
      \While{All data features}
        \State $OneHotEndcode(x).$
        \State $Normalize(x).$
        \State Encode $|\psi(x)\rangle \leftarrow x$
      \EndWhile
      \State \textbf{return} $|\psi(x)\rangle$.
    \EndProcedure

    \Procedure{orb-QFL}{\textit{Train Data ($|\psi(x)\rangle$, y)}}
    \State Initial weight $\theta_s$ = None
      \For{$r$ in range(R)}
        \For{$satNode, i$ in $satList[]$}
          \If{$r == 0$ and $i == 0$}
           \State Initialize ${\theta_{satNode}^{VQC}}$
            \State $VQC_i.fit(\theta_i, |\psi(x)\rangle,y)$
            \State Save $\theta_s = satNode_{VQC}^{\theta_i}$
            \State Compute $dist(satNode_i, sateNode_{(i+1)})$
            \State Transmit($fileS(\theta), dist, delay$)  
          \Else
          \State Receive model from $SatNode_{(i-1)}$
            \State Load $VQC_i.\theta_i = \theta_s$
            \State $VQC.classifier.warm\_start = True$
            \State $VQC_i.fit(\theta_s, |\psi(x)\rangle,y)$
           \State Save $\theta_s = satNode_{VQC}^{\theta_i}$
           \State Compute $dist(satNode_i, sateNode_{(i+1)})$
           \State Transmit($fileS(\theta), dist, delay$)  
          \EndIf
        \EndFor
      \EndFor
    \EndProcedure
  \end{algorithmic}
\end{algorithm}}

\section{Theoretical Analysis}
In this section, we present theoretical analysis, examine different aspects, and discuss convergence.

\begin{assumption}[Lipschitz Continuity]
The objective functions for COBYLA maintain the Lipschitz continuity. This assumption ensures that the linear approximations employed by COBYLA remain precise within the trust region, contributing to a more reliable and quicker convergence.
\end{assumption}

\begin{assumption}[Uniform $L$-smoothness]
The functions $\{ F_1, \dots, F_N \}$ are uniformly $L$-smooth for $L \geq 0$. For arbitrary vectors $x$ and $y$, the following inequality is satisfied:
    \[F_k(y) \leq F_k(x) + \langle y - x, \nabla F_k(x) \rangle + \frac{L}{2} \|y - x\|^2\]
for all $k = \{1, 2, \dots, N \}$, which serves as a quadratic upper bound for an $L$-smooth function.
\end{assumption}

\begin{assumption}[$\mu$-strong Convexity]
Functions \(\{ F_1, \cdots, F_N \}\) are all \(\mu\)-strongly convex: for all \( x \) and \( y \),
\[   F_k(x) \geq F_k(y) + (x - y)^T \nabla F_k(y) + \frac{\mu}{2} \|x - y\|_2^2.
\]
\end{assumption}

\begin{assumption}[Data Distribution Similarity]
Consider a set of \(n\) clients participating, denoted as \(\{C_1, C_2, \dots, C_K\}\). We assume that the data distributions of the data sets of these clients, represented by probability distributions \{\(P_1, P_2, \dots, P_K\)\}, exhibit a degree of similarity such that
\[    \forall i, j \in \{1, 2, \dots, K\}, \quad \delta(P_i, P_j) \leq \epsilon.\]
\(\delta(\cdot, \cdot)\) denotes a divergence measure (e.g., total variation distance, Kullback-Leibler divergence) between the distributions and \(\epsilon\) is a small positive constant representing the maximum divergence allowed between any pair of distributions.
\end{assumption}

\begin{assumption}[Device and Environmental Conditions]
Some assumptions we consider in terms of device and environmental conditions are as follows:
\begin{compactenum}
    \item The devices are fully functional during local training without any failure.
    \item Environmental factors such as rain, temperature, ions, etc., are considered to be ideal for the experimental simulation.
    \item For experimental purposes, we assume that line of sight is visible for successful communication between satellite and ground stations (GS) or between satellite to satellite, i.e., after each device training, the device immediately sends the model to another device without any issue.
\end{compactenum}
\end{assumption}
\begin{lemma}
Let \( F \) be a continuous Lipschitz function with Lipschitz constant \( L \). The regret \( R(T) \) of the COBYLA optimizer  
after \( T \) iterations is limited by \cite{Powell1994}
   \[R_F(T) = \sum_{k=1}^{T} [ F(\theta_t) - F(\theta^*) ] \leq L \sum_{t=1}^{T} \Delta_t\]
where,  
\( \theta_t \) is the position at iteration \( t \), \( \theta^* \) is the best solution, and \( \Delta_t \) is the trust region radius at iteration \( t \).
\end{lemma}

The error limit at iteration \( t \) can be explained in terms of the trust region radius \( \Delta_t \) and the precision of the linear approximations. 
With the bound on the trust region radius,
\[\| \theta_{t+1} - \theta^* \| \leq \Delta_t\]
where \( \theta^* \) is the best solution and \( \Delta_t \) is the trust region radius at iteration \( t \). 
With a bound on the objective function error,
\[| f(\theta_{t+1}) - f(\theta^*) | \leq C \Delta_t\]
where \( C \) is a constant dependent on the Lipschitz continuity of the objective function \( F \) and the accuracy.

\begin{theorem}
    Let Assumptions 1 to 5 hold, with strongly convex local objectives, 
 $\Delta_t$ trust region radius adjustments at time $t$, 
the expected difference between the value of the function in the model \(\theta^{R}\) and the optimal value of the function \(F(\theta^*)\) is bounded by \cite{liConvergenceAnalysisSequential}
\begin{align}
E\left[F(\theta^{R}) - F(\theta^*)\right] \notag 
&\leq \underbrace{L \sum_{t=1}^{T} \Delta_t}_{\text{Optimizer}} \\
&+ \underbrace{\mu (\theta_0 - \theta^*)^2 \exp\left(-\frac{\mu \Delta_t R}{2}\right)}_{\text{FL}} \notag \quad \\
&+ \frac{\Delta_t }{NK} + \frac{L \Delta_t^2 }{NK} \\
&+ \underbrace{\gamma_c \tau_c R + \delta_c \rho_{loss} \rho 
+ \epsilon_c \frac{\rho}{B} T}_{\text{SATCOM Factors}} \notag \quad \\
&+ \underbrace{\alpha_q \sigma_q^2 N_q}_{\text{Quantum Factor}} \notag
\end{align}

\end{theorem}
where, 
\(\theta^{R}\) is the model parameters after rounds \(R\) for $N$ satellites with $K$ local updates, \(F(\theta)\) is the global objective function, \(\theta^*\) is the optimal solution of the global objective function, 
In addition, 
    $\gamma_c \tau_c R$ is impact of communication latency, where $\gamma_c$ is a constant, $\tau_c$ is the latency;
    $\delta_c \rho_{loss} \rho$ is impact of packet loss, where $\delta_c$ is a constant, $\rho_{loss}$ is the packet loss rate, and $\rho$ is the amount of data transmitted;
    $\epsilon_c \frac{\rho}{B} T$ is impact of bandwidth constraints, where $\epsilon_c$ is a constant, $\rho$ is the amount of data transmitted, $B$ is the bandwidth and $T$ is the duration of communication.
Whereas, 
    $\alpha_q \sigma_q^2 N_q$ is impact of quantum noise, where $\alpha_q$ is a constant, $\sigma_q$ is the noise level, and $N_q$ is the number of qubits.
The proofs are provided in the appendix \ref{sec:appendix}.

\section{Experiment}
\subsection{Set up \& Parameters.}

\subsubsection*{\textbf{LEO Constellation.}}
We experiment with $5$ and $10$ satellites in an LEO constellation that are equidistant from each other, covering the entire Earth, each in its orbit around the Earth in two different sets of experiments.
Each orbit of the satellite is located at a height of $500$ km above the Earth's surface, with an inclination angle of $60$ deg and an angular spacing of $360/n$.
For only a QFL setting, a geostationary satellite is considered which acts as a server.
For which, we experimented with $2$ settings: the first geostationary satellite was located on Earth at about 20 m altitude; the second geostationary satellite is an actual GEO satellite with an altitude of $35786 * u.km$.

\subsubsection*{\textbf{Tools.}}
\textit{Matlab satellite communications toolbox} is used to simulate, analyze, and test satellite communications systems and links.
For performance simulation, the \textit{Qiskit framework} is used, while for satellite simulation, the \textit{Poliastro library} is used.
Poliastro is a Python-based library that can be used for interactive astrodynamics and orbital mechanics.
Thus, we used the library to create satellite orbit paths and track their propagation, through which we keep a record of the distance between the satellites.
Qiskit is an open-source SDK for working with quantum computers at the levels of quantum circuits, operators, and primitives.

\subsubsection*{\textbf{Dataset Used.}}
For our experimental simulation, we used the Statlog (Landsat Satellite) dataset
\cite{misc_statlog_(landsat_satellite)_146}.
Statlog data are a multivariate data set that consists of multispectral values of pixels in $3$ x $3$ neighborhoods in satellite images.
In total, it has $6435$ samples with 36 features and $7$ labels from $1$ to $7$ where each number relates to red soil, cotton crop, grey soil, damp gray soil, soil with vegetation stubble, mixture class and very damp grey soil, respectively.
Figure \ref{fig:statlog_dataset}, shows the statlog datasets after reducing the dimension of the feature.
$90\%$ of the dataset is used for training, while the rest is used for testing.
The data set is also distributed among varied numbers of devices in different sets of experiments of $10$ or $5$ devices.

\subsection{Results.}

\begin{figure}[!htb]
    \centering
     \begin{subfigure}[b]{0.45\columnwidth}
        \centering
        \includegraphics[width=\columnwidth]{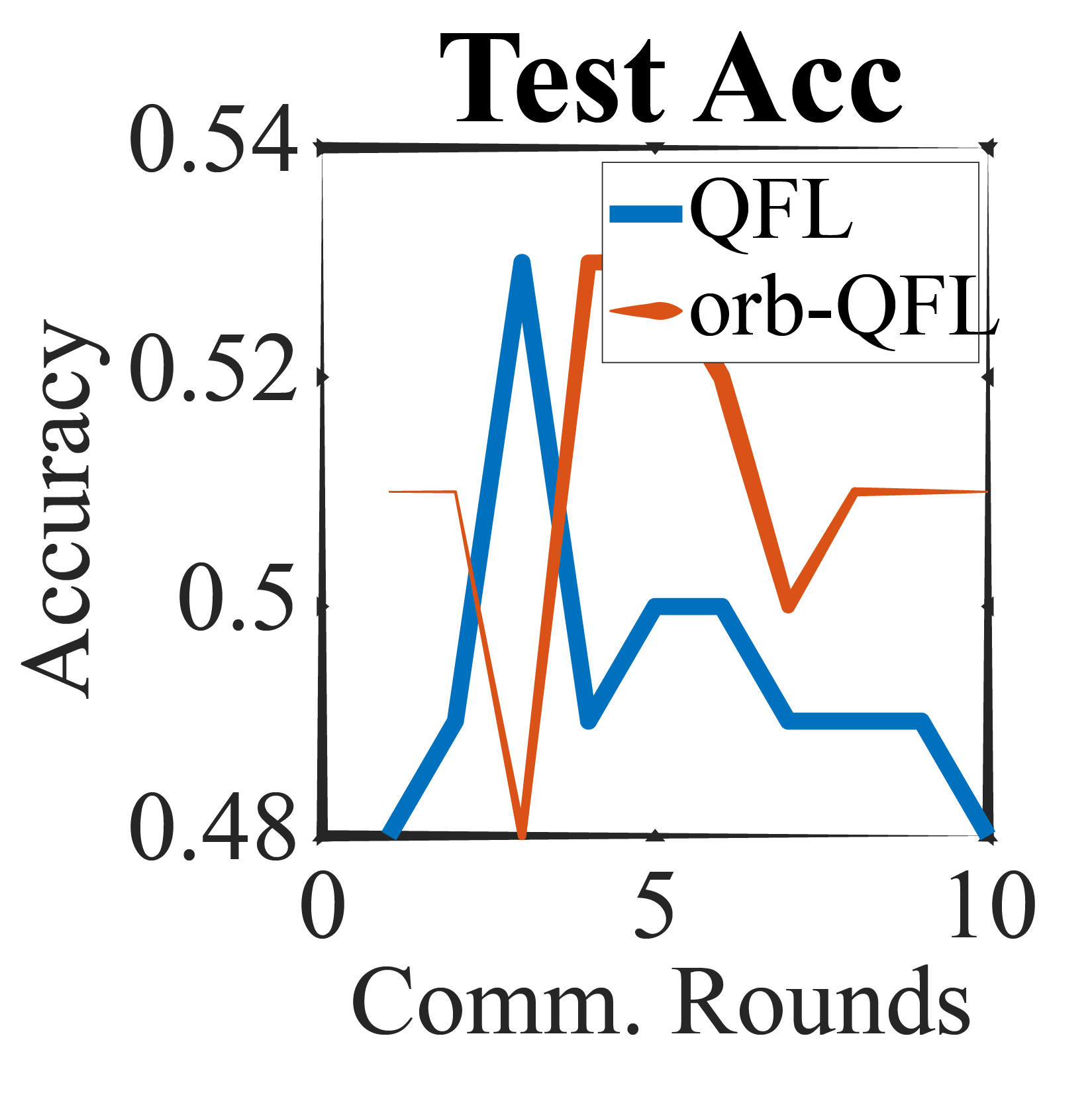}
        \caption{Test Acc}
        \label{fig:server_test_acc}
    \end{subfigure}
    \begin{subfigure}[b]{0.45\columnwidth}
        \centering
        \includegraphics[width=\columnwidth]{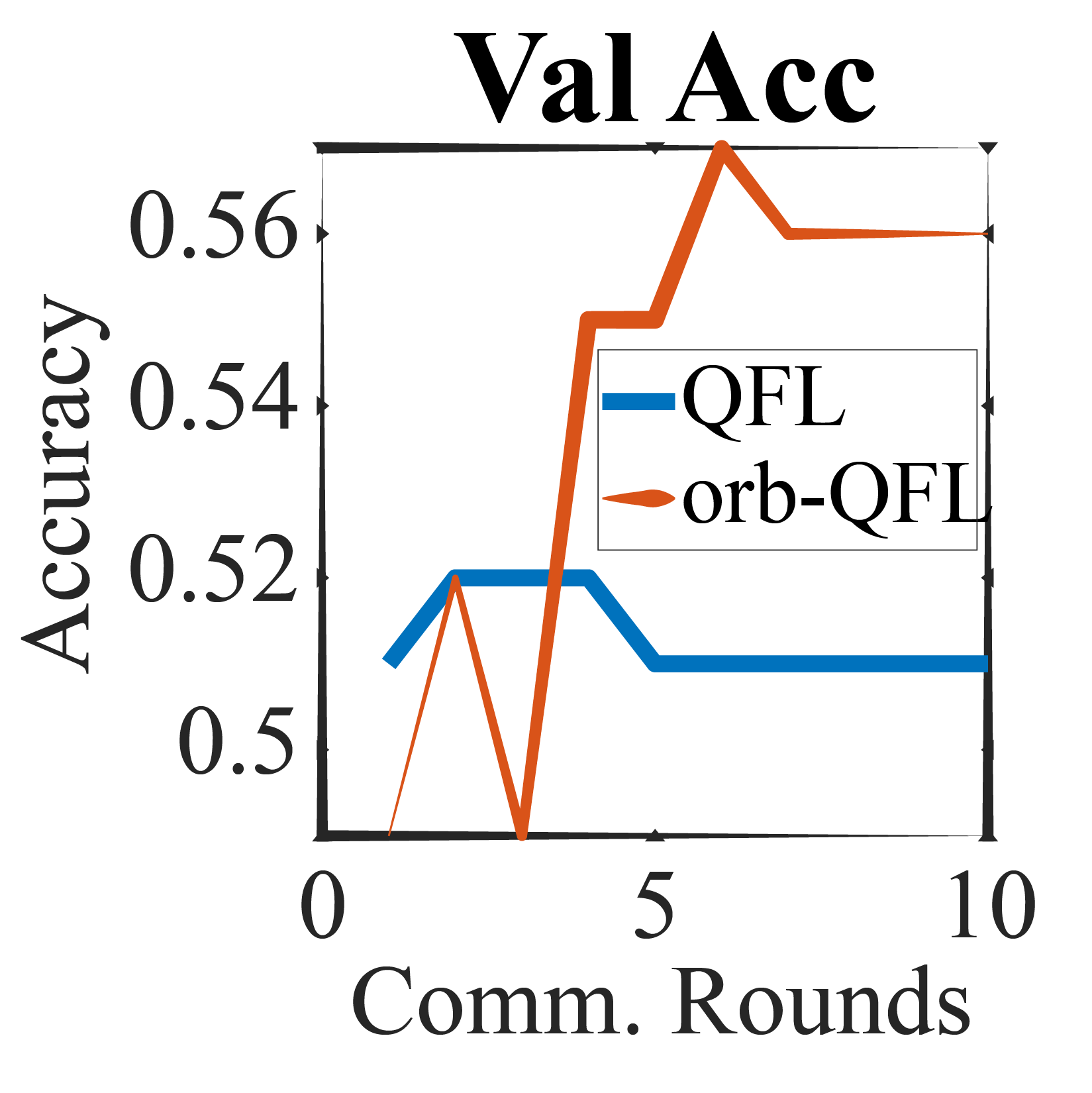}
        \caption{Val Acc}
        \label{fig:server_train_acc}
    \end{subfigure}
    \caption{Server Performance}
    \label{fig:performance_test_accuracy}
\end{figure}

\textbf{Accuracy.}
To evaluate the performance of the model at the end of each communication round, a hypothetical server is added to orb-QFL only for testing purposes.
From Figures \ref{fig:server_test_acc} and \ref{fig:server_train_acc}, we can see that orb-QFL performs better in terms of server performance.
This implies that the proposed method is better suited in terms of both training and test performance.
It is promising to see better performance considering we have omitted the need for server and server-related communications and activities like aggregation etc.

\begin{figure}
     \centering
    \begin{subfigure}[b]{0.45\columnwidth}
        \centering
        \includegraphics[width=\columnwidth]{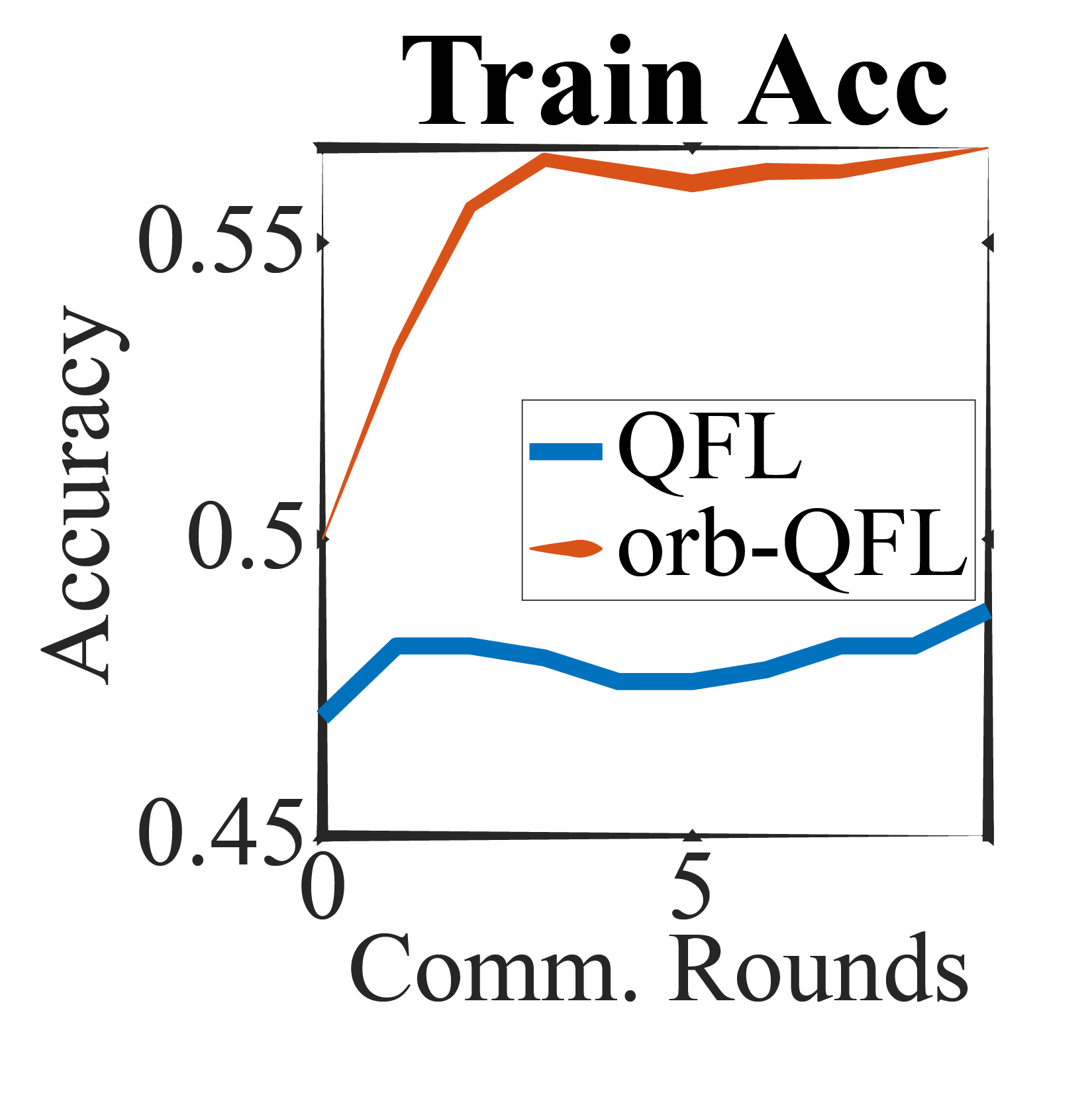}
         \caption{Avg Train Acc}
        \label{fig:devices_train_acc_avg}
    \end{subfigure}
    \begin{subfigure}[b]{0.45\columnwidth}
        \centering
        \includegraphics[width=\columnwidth]{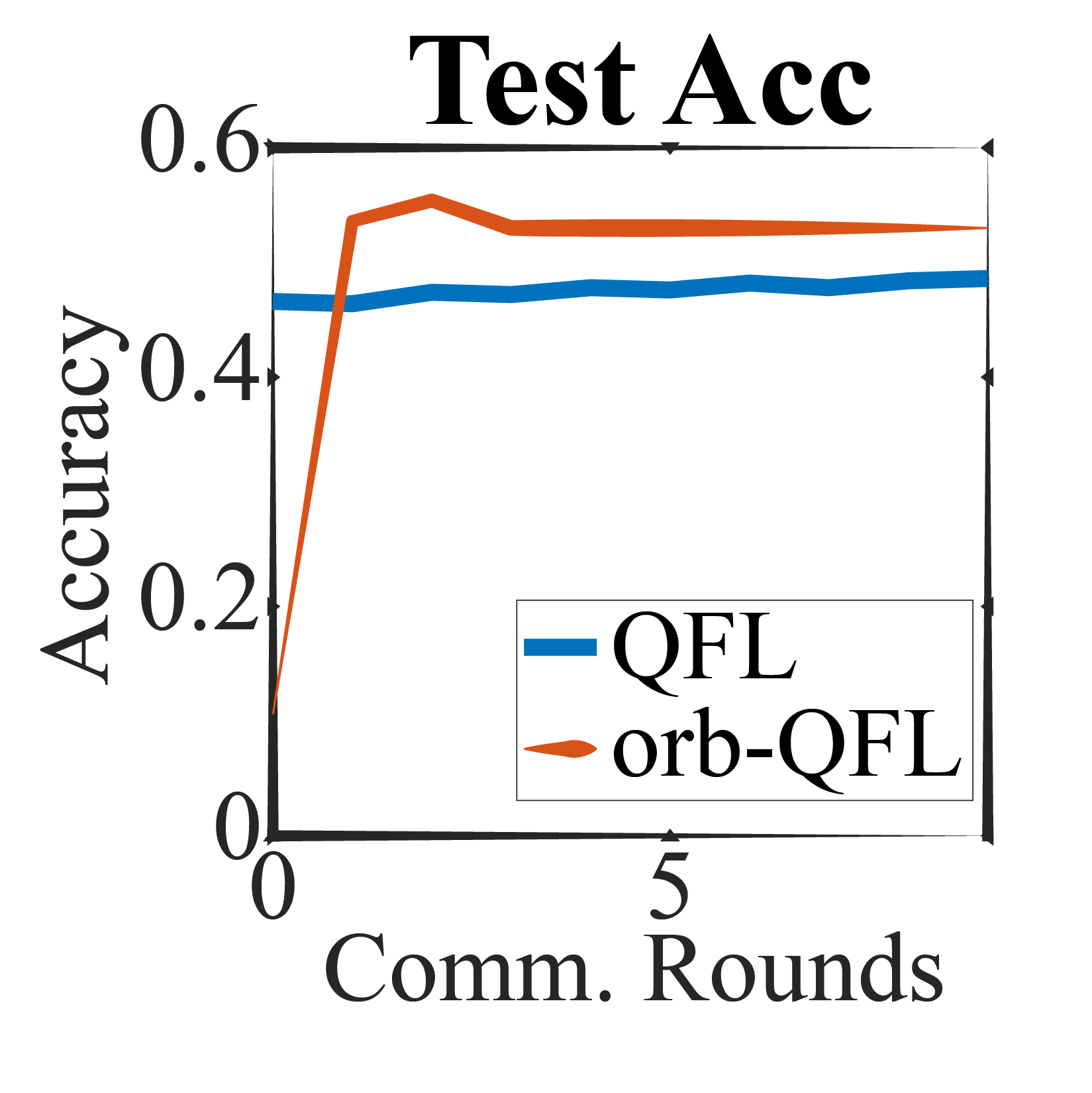}
         \caption{Avg Test Acc}
        \label{fig:devices_test_acc_avg1}
    \end{subfigure}
    \caption{Device Performance}
    \label{fig:performance_device}
\end{figure}

Similar results can be seen in Figures \ref{fig:devices_train_acc_avg} and \ref{fig:devices_test_acc_avg1} in terms of devices' performance.
The overall average accuracy for all devices in terms of training is consistently better with orb-QFL as shown in Figure \ref{fig:devices_train_acc_avg}. While in testing, the performance is again slightly improved with orb-QFL as shown in Figure \ref{fig:devices_test_acc_avg1}.

\begin{figure}
    \centering
    \begin{subfigure}[b]{0.45\columnwidth}
        \centering
        \includegraphics[width=\columnwidth]{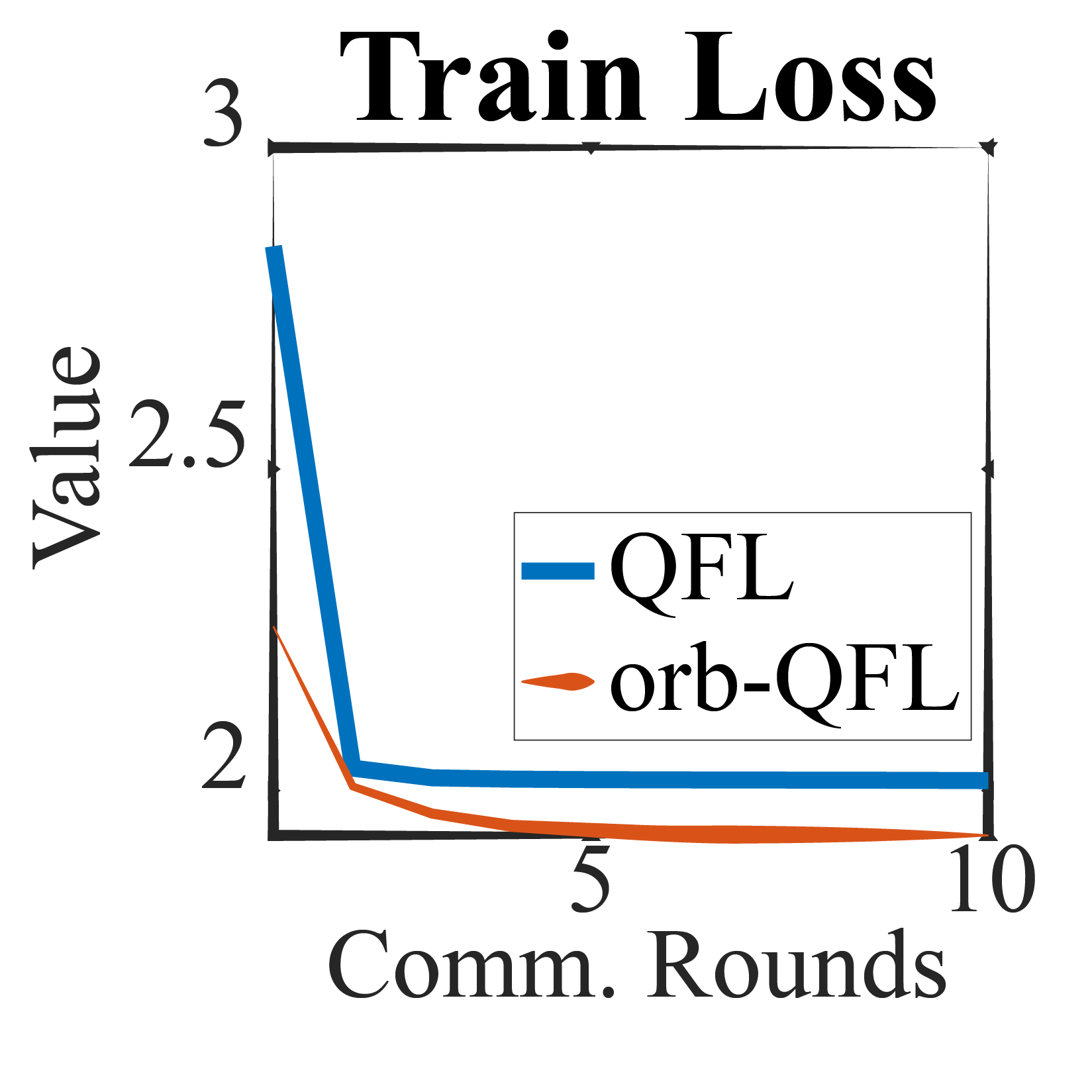}
        \caption{Avg Devices Loss}
        \label{fig:devices_avg_loss}
    \end{subfigure}
     \begin{subfigure}[b]{0.45\columnwidth}
        \centering
        \includegraphics[width=\columnwidth]{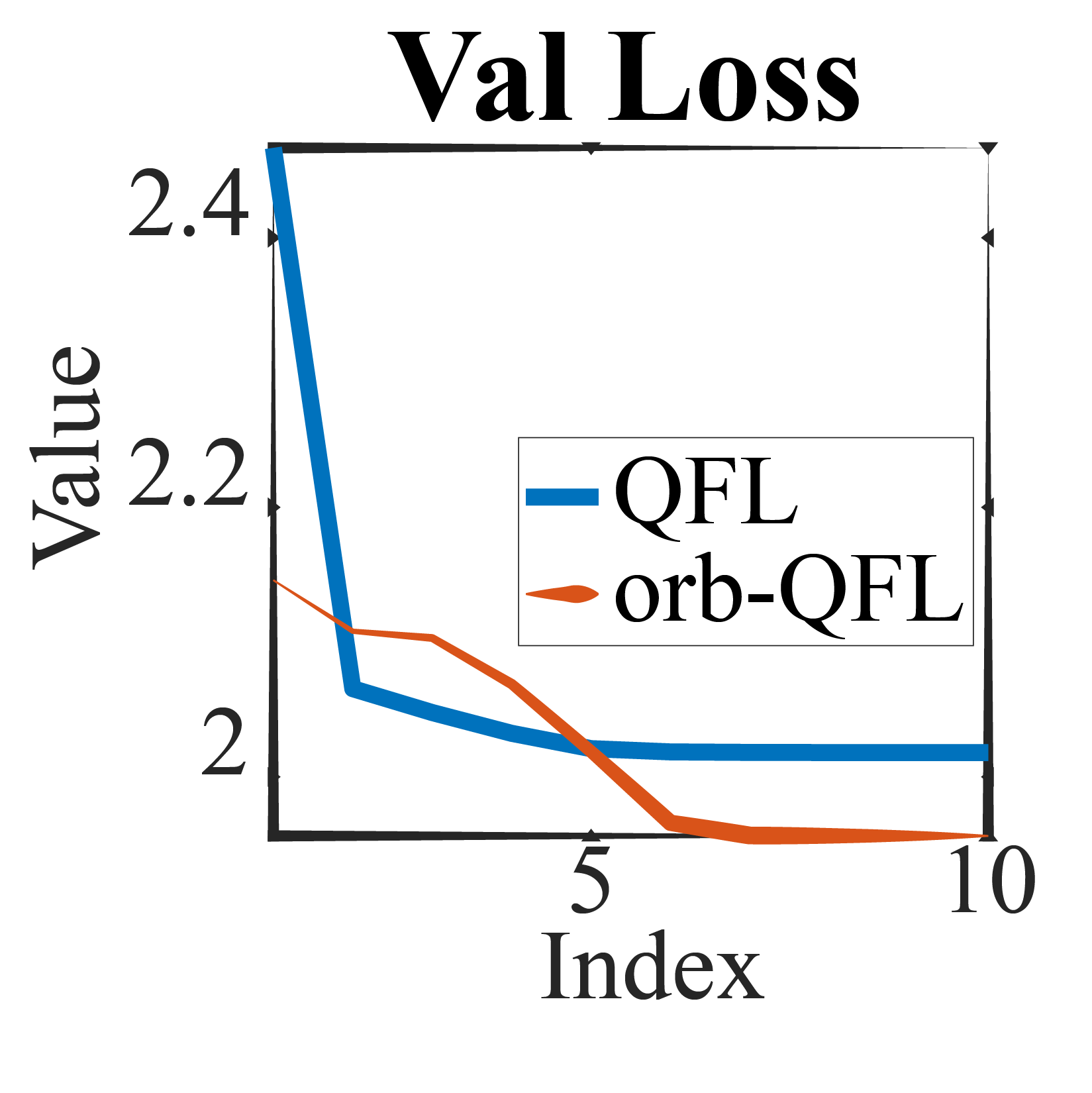}
        \caption{Server Val Loss}
        \label{fig:server_loss}
    \end{subfigure}
    \caption{Objective Values}
    \label{fig:performance}
\end{figure}

\textbf{Loss.}
With objective functions, we understand the loss value to examine how the optimizer performs and converges during the optimization process.
With a maximum value of 100 for the COBYLA optimizer, the convergence is studied with experimental analysis.

The result for the average of all devices is shown in Figure \ref{fig:devices_avg_loss}.
Similarly, for the server, the objective values are better with orb-QFL, which is shown in Figure \ref{fig:server_loss}.
These results point towards better practicality and performance of the proposed approach.

\textbf{Link Budget Analysis.}
Link Budget Analysis was performed using Matlab.
We consider two LEO satellites at an altitude of 500 km, which are separate by 72 degrees in terms of longitude. A geostationary satellite is considered with altitude of 20 m above earth surface almost in the equidistant from both satellites. For this experiment, we neglected the loss factors from all satellites and the geostationary satellite.
Thus, we compare the links between station to satellite (L1), satellite to station (L2) , and satellite to satellite (L3).
The properties of links (L1, L2) include a frequency of 2 GHz, a bandwidth of 6 MHz, a bitrate of 10 Mbps, and a required Eb/No of 10 dB. The transmitter on Satellite (S1) has a Tx HPA power of 17 dBW, a Tx HPA OBO of 6 dB, and a Tx antenna gain of 60 dBi. The receiver on Satellite (S2) has an Rx G/T of 10 dB/K. Additionally, the properties of Link (L3) include a frequency of 2.2 GHz, a bandwidth of 5 MHz, a bitrate of 10 Mbps and a required Eb/No of 10 dB.

\begin{figure}[!htbp]
    \centering
    \begin{subfigure}[b]{0.45\columnwidth}
        \centering
        \includegraphics[width=\linewidth]{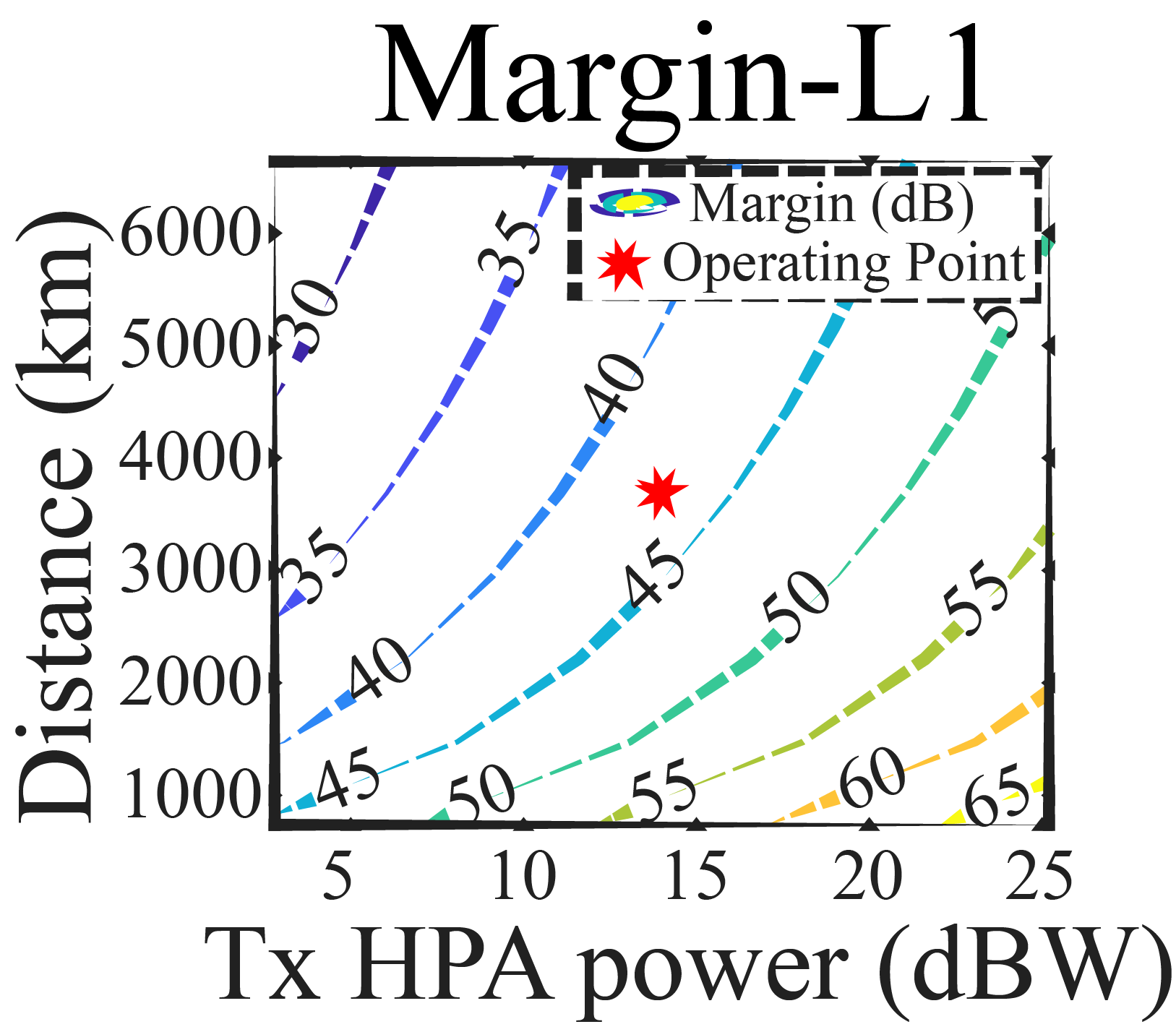}
        \caption{S2G}
        \label{fig:s2g}
    \end{subfigure}
    \begin{subfigure}[b]{0.45\columnwidth}
        \centering
        \includegraphics[width=\linewidth]{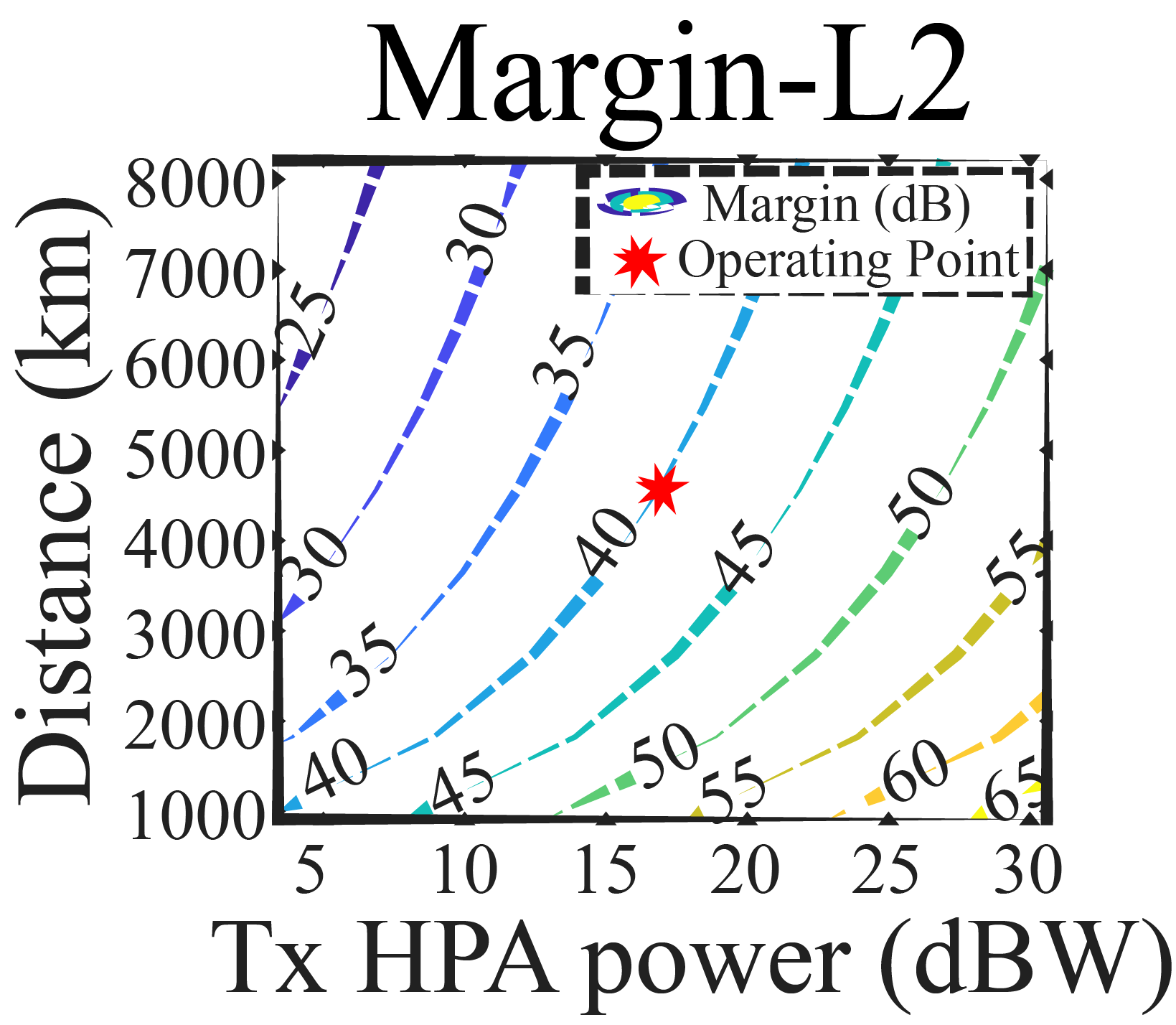}
        \caption{G2S}
        \label{fig:g2s}
    \end{subfigure}
    \begin{subfigure}[b]{0.45\columnwidth}
        \centering
        \includegraphics[width=\linewidth]{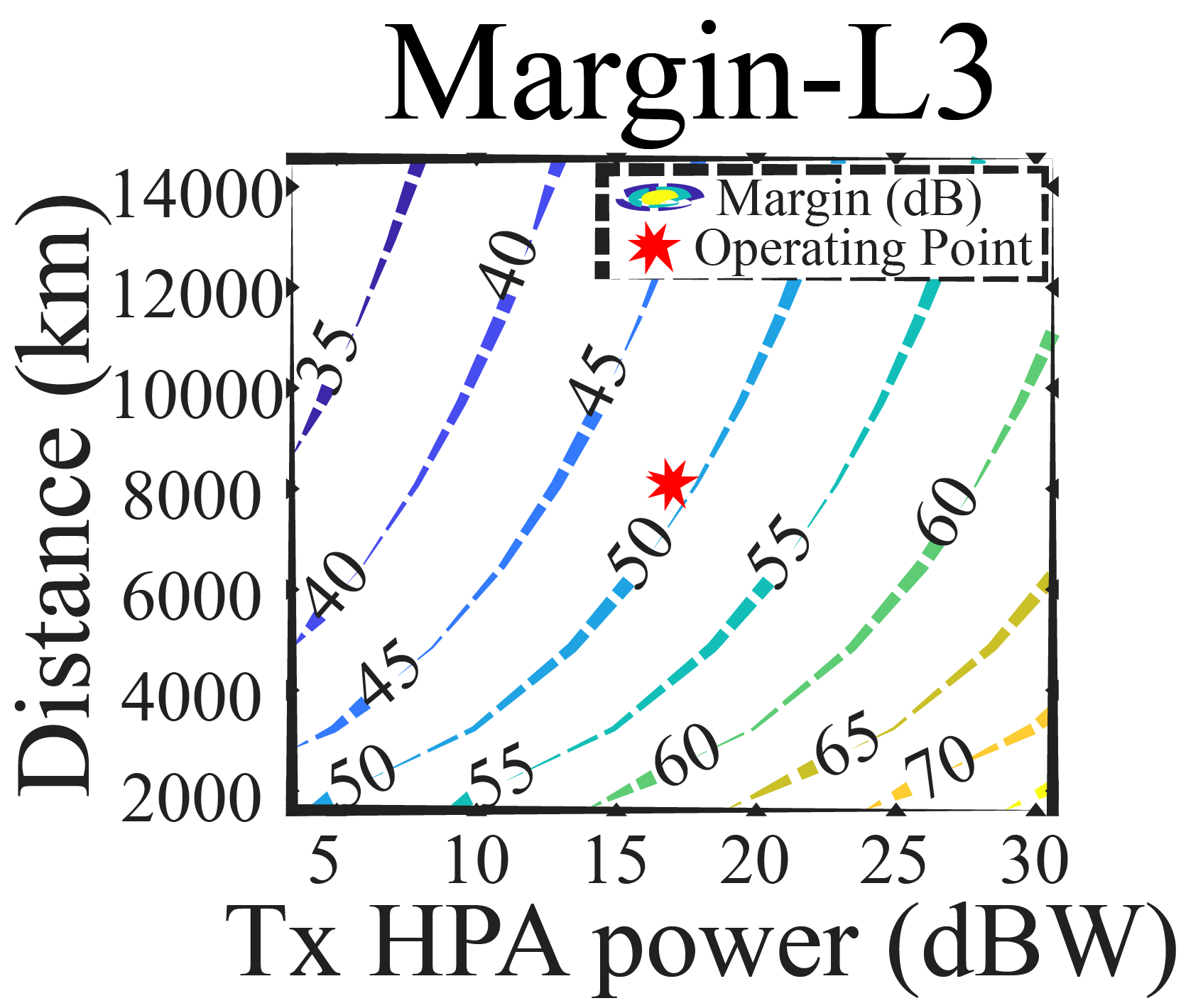}
        \caption{S2S}
        \label{fig:s2s}
    \end{subfigure}
    \begin{subfigure}[b]{0.45\columnwidth}
    \includegraphics[width=\linewidth]{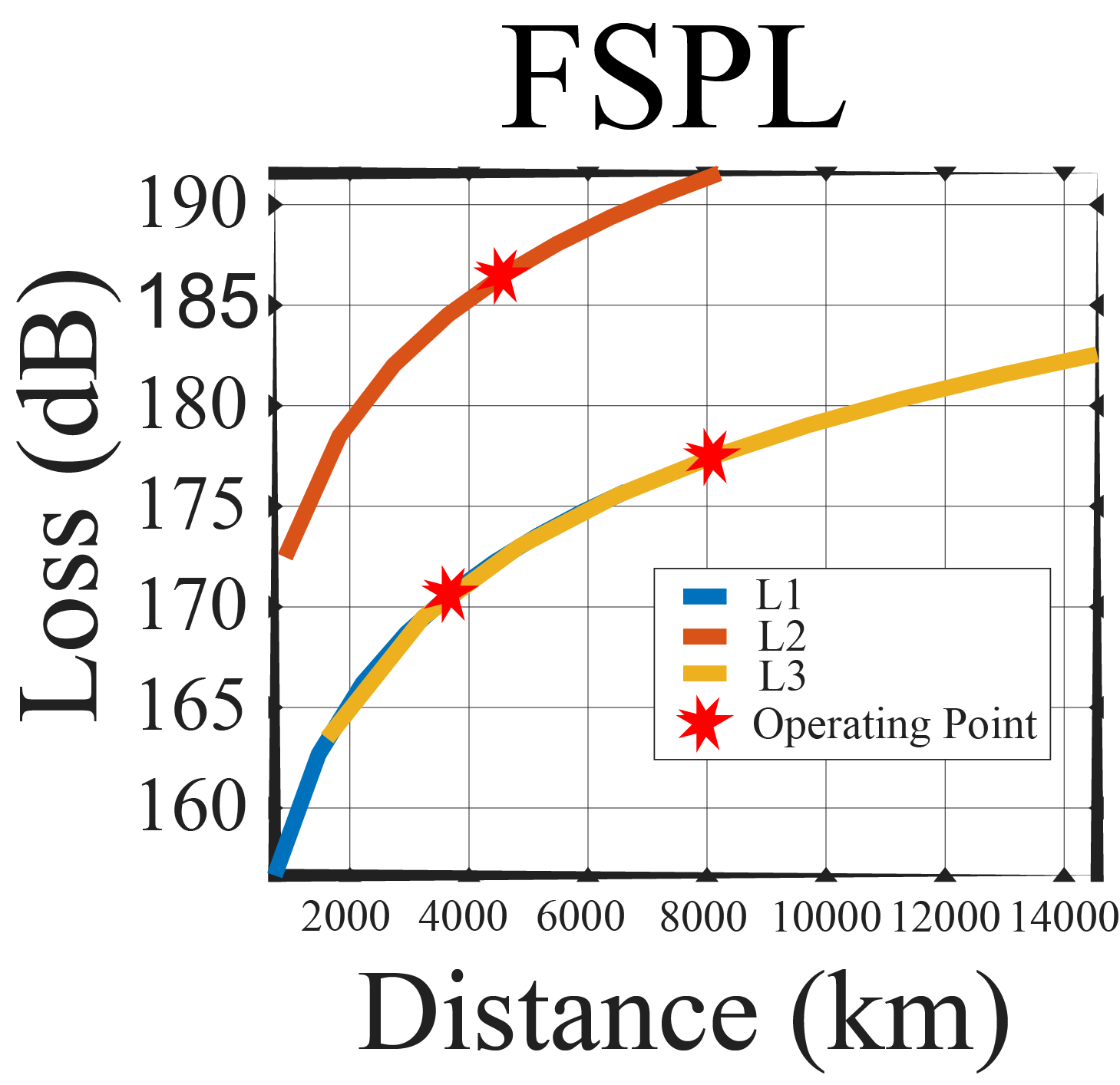}
    \caption{FSPL}
    \label{fig:fspl_vs_distance}
    \end{subfigure}
    \begin{subfigure}[b]{0.45\columnwidth}
    \includegraphics[width=\linewidth]{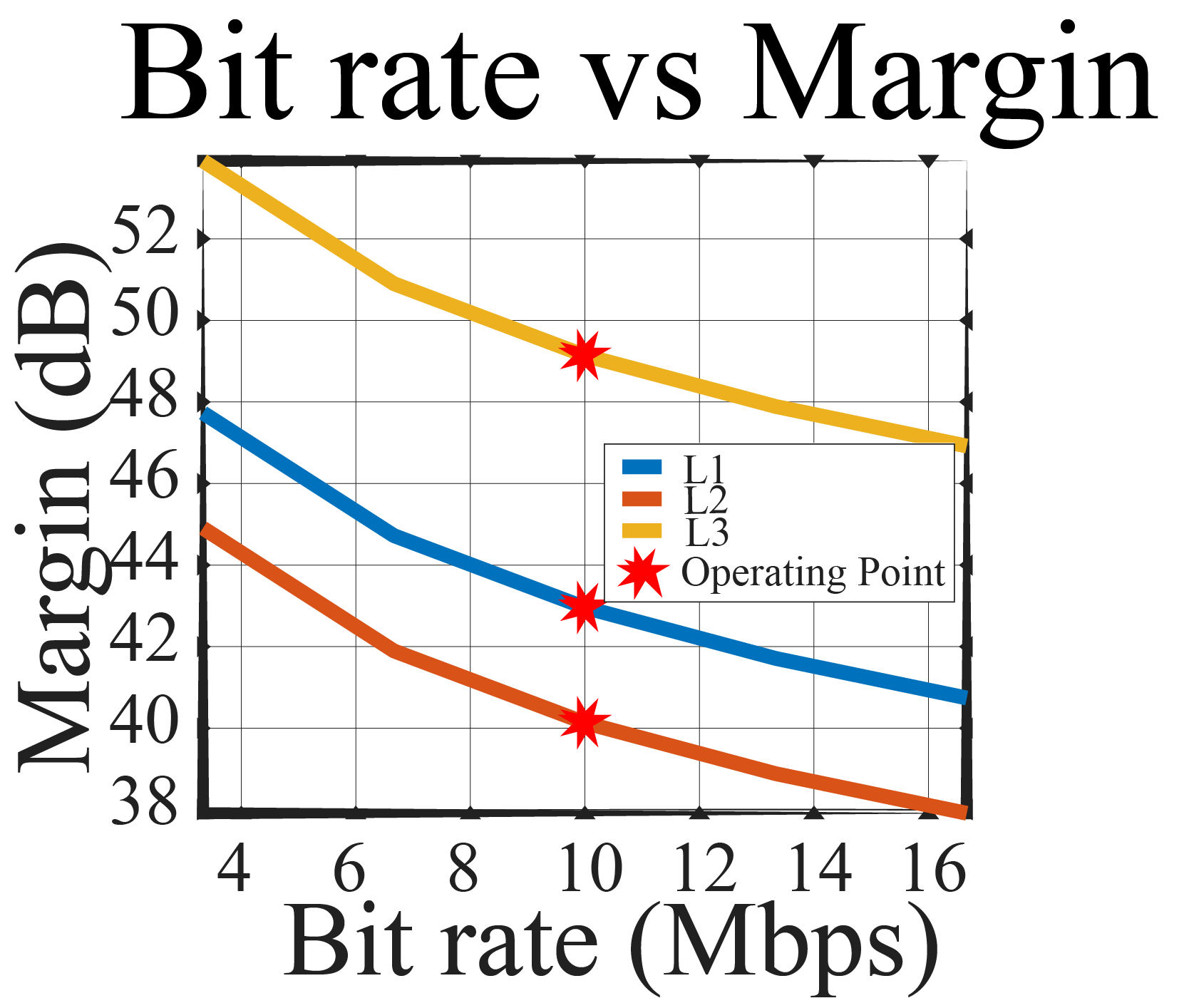}
    \caption{Bitrate}
    \label{fig:margin_vs_bitrate}
    \end{subfigure}
    \caption{Link Budged Analysis}
    \label{fig:link_budged_analysis}
\end{figure}

From the contour plot shown in Figures \ref{fig:s2g}, \ref{fig:g2s} and \ref{fig:s2s}, we compare the high amplifier power (HPA) versus distance (km). 
With contour lines, we observe the margin value (dB), which is the excess SNR threshold value required for reliable communication. 
From Figures \ref{fig:s2g}, \ref{fig:g2s} and \ref{fig:s2s}, G2S operates at around 44 dB and 14 dBW power, S2G at around 40 dB with 17 dbW power and the
S2S link at about 50 dB with 16 dbW power. 
The inter-satellite communication clearly has a better margin, which indicates that communication to and from the Earth station or any server thereby needs more power and provides less margin which can cause unreliable communication in comparison to orb-QFL setting where most transfer and receiving happens only between satellites themselves.

 Figure \ref{fig:fspl_vs_distance} represents Free Space Path Loss (FSPL) versus distance for three different links L1, L2 and L3.
In this case, L1 is the best-performing link in terms of FSPL having the lowest path loss at a given distance which means it experiences the least signal degradation over distance.
Whereas, L2 suffers from the highest FSPL.
Thus, L3 comes in between two other link properties.

   

Figure \ref{fig:margin_vs_bitrate} shows the margin (in dB) as a function of the bit rate (in Mbps) for the links.
All lines display a decrease in margin with an increase in bitrate which is as expected. 
Amongst the links, L3 has the highest margin than the others indicating better reliability than the others.
Thus, orb-QFL which is dependent more on direct satellite-to-satellite communication provides further advantages along with improved performance of the QFL network.
    

Lastly, it is to be noted that the default QFL means that both L1 and L2 links are involved.
But in orb-QFL, there is only an L3 link available with only need for L1 and L2 link communication once a while only after the end of all the communication rounds.
Thus, these results show how robust and reliable orb-QFL can be validating the practicality of this work.

\section{Conclusion}
In this work, we first presented a novel approach to continuous QFL framework integration training without any server aggregation procedure involved.
Along with it, our study explores the applicability of QFL with our approach to LEO satellite network communications to improve communication efficiency and solve many bottleneck issues involved in SATCOM settings. 
Through our theoretical and experimental investigation, we demonstrated both the practicality and feasibility of our method within satellite networks.
Our research will serve as an essential foundation for future studies in quantum federated learning (QFL) and satellite constellation communications, paving the way for advancements in next-generation machine learning and communication networks.


\section{APPENDIX} \label{sec:appendix}
\subsection{\textbf{Proof of Lemma 1.}}
Since \( F \) is Lipschitz continuous with constant \( L \), for any \( \theta \) and \( \theta' \),
\[|F(\theta) - F(\theta')| \leq L \|\theta - \theta'\|.\]
COBYLA optimizer constructs a trust region around the current point \( \theta_t \) and solves a subproblem within this region
\[\min_{d} F(\theta_t + d)\]
subject to
$\|d\| \leq \Delta_t$.
Here, \( \Delta_t \) is the radius of the trust region.

In each iteration \( t \), COBYLA updates the point \( \theta_t \) to \( \theta_{t+1} \) by solving the sub-problem, ensuring that \( \|\theta_{t+1} - \theta_t\| \leq \Delta_t \).
The regret \( R_F(T) \) is the sum of differences between the function values at \( \theta_t \) and \( \theta^* \)
\[
R_F(T) = \sum_{t=1}^{T} [ F(\theta_t) - F(\theta^*) ].
\]

Using the Lipschitz continuity of \( F \), we can bound the difference as, 
\[F(\theta_t) - F(\theta^*) \leq L \|\theta_t - \theta^*\|.\]
By triangle inequality, we know
\[\|\theta_t - \theta^*\| \leq \|\theta_t - \theta_{t-1}\| + \|\theta_{t-1} - \theta^*\| 
\]
Since $ \|\theta_{t+1} - \theta_t\| \leq \Delta_t $,  by iterating the inequality, we can have, 
\[\|\theta_t - \theta^*\| \leq \sum_{k=1}^{t} \Delta_k.\]
Thus,
\[
F(\theta_t) - F(\theta^*) \leq L \|\theta_t - \theta^*\| \leq L \sum_{k=1}^{t} \Delta_k.
\]
Summing the bound on $F(\theta_t) - F(\theta^*)$ over \( t \) from 1 to \( T \),
\[
R_F(T) = \sum_{t=1}^{T} [ F(\theta_t) - F(\theta^*) ] \leq L \sum_{t=1}^{T} \Delta_t.
\]

This proof confirms that the regret \( R_F(T) \) of the COBYLA optimizer after \( T \) iterations is bounded by \( L \sum_{t=1}^{T} \Delta_t \).

\subsection{\textbf{Proof of Theorem 1.}}
From \cite{liConvergenceAnalysisSequential}, we have upper bound for global model with learning rate $\eta$, effective learning rate denoted by \(\tilde{\eta}\)  with Stochastic Gradient Descent as, 
\begin{align}
E\left[F(\theta^{R}) - F(\theta^*)\right] \leq 
&\frac{9}{2} \mu D^2 \exp\left(-\frac{\mu \tilde{\eta} R}{2}\right) \notag \\
&+ \frac{12 \tilde{\eta} \sigma^2}{NK} + \frac{18L \tilde{\eta}^2 \sigma^2}{NK} + \frac{18L \tilde{\eta}^2 \zeta^2_*}{N} \notag
\end{align}

where,  \(\theta^{R}\) denotes the weighted average of the global model parameters after \(R\) rounds. \(F(\theta)\) represents the global objective function and \(\theta^*\) signifies the optimal solution to this global objective function. The effective learning rate is denoted by \(\tilde{\eta}\).
\(D\) indicates the initial distance between the starting point and the optimal solution. The variance of the stochastic gradient is represented by \(\sigma^2\).
\(\zeta^2_*\) is a constant that measures the heterogeneity of local objective functions in the optimal solution.

Considering the optimizer factor, we have, 
\begin{align}
E\left[F(\theta^{R}) - F(\theta^*)\right] 
&\leq L \sum_{t=1}^{T} \Delta_t \notag\\
&+ \frac{9}{2} \mu D^2 \exp\left(-\frac{\mu \tilde{\eta} R}{2}\right) \notag \\
&+ \frac{12 \tilde{\eta} \sigma^2}{NK}
+ \frac{18L \tilde{\eta}^2 \sigma^2}{NK} \notag\\
&+ \frac{18L \tilde{\eta}^2 \zeta^2_*}{N} \notag
\end{align}
Then, 
\begin{align}
E\left[F(\theta^{R}) - F(\theta^*)\right] 
&\leq \frac{L}{R} \sum_{t=1}^{R} \Delta_t \notag\\
&+ \frac{9}{2} \mu D^2 \exp\left(-\frac{\mu \tilde{\eta} R}{2}\right) \notag \\
&+ \frac{12 \tilde{\eta} \sigma^2}{NK} 
+ \frac{18L \tilde{\eta}^2 \sigma^2}{NK} \notag\\
&+ \frac{18L \tilde{\eta}^2 \zeta^2_*}{N} \notag
\end{align}
Replacing $\tilde{\eta}$ with $\Delta_t$ and relaxing heterogeneity aspect, 
\begin{align}
&\leq \frac{L}{R} \sum_{t=1}^{R} \Delta_t 
+ \frac{9}{2} \mu D^2 \exp\left(-\frac{\mu \Delta_t R}{2}\right) \notag \\
&+ \frac{12 \Delta_t \sigma^2}{NK} + \frac{18L \Delta_t^2 \sigma^2}{NK} \notag \\
\notag
\end{align}

Now considering, Sat-Comm and Quantum computational factors, 
\begin{align}
&\leq \frac{L}{R} \sum_{t=1}^{R} \Delta_t \\
&+ \frac{9}{2} \mu D^2 \exp\left(-\frac{\mu \Delta_t R}{2}\right) \\
&+ \frac{12 \Delta_t \sigma^2}{NK} \notag 
+ \frac{18L \Delta_t^2 \sigma^2}{NK}  \\
&+\Delta_{C} +\Delta_{Q}
\end{align}
where, 
$\Delta_{C} = \gamma_c \tau_c R + \delta_c \rho_{loss} \rho + \epsilon_c \frac{\rho}{B} T$
and 
$\Delta_{Q} = \alpha_q \sigma_q^2 N_q$.

Finally, with further simplification, 
\begin{align}
E[F(\bar{x}_R) - F(x^*)] 
&\leq \frac{L}{R} \sum_{t=1}^{R} \Delta_t \\
&+ \mu D^2 \exp \left( -\frac{\mu \Delta_t R}{2} \right) \\
&+ \frac{\Delta_t }{NK}
+ \frac{L \Delta_t^2 }{NK} + \Delta_{C}  + \Delta_{Q}
\end{align}
Then, 
\begin{align}
& E\left[F(\theta^{R}) - F(\theta^*)\right] \notag\\
&\leq \underbrace{L \sum_{t=1}^{T} \Delta_t}_{\text{Optimizer}} \\
&+ \underbrace{\mu (\theta_0 - \theta^*)^2 \exp\left(-\frac{\mu \Delta_t R}{2}\right)}_{\text{FL}} \notag\\
&\quad + \frac{\Delta_t }{NK} + \frac{L \Delta_t^2 }{NK} \notag \quad \\
&+ \underbrace{\gamma_c \tau_c R + \delta_c \rho_{loss} \rho
+ \epsilon_c \frac{\rho}{B} T}_{\text{SATCOM Factors}} \\
&+ \underbrace{\alpha_q \sigma_q^2 N_q}_{\text{Quantum Factor}} \notag.
\end{align}

\subsection{\textbf{Statlog Dataset.}}
\begin{figure}[!htb]
    \centering
    \includegraphics[width=0.8\columnwidth]{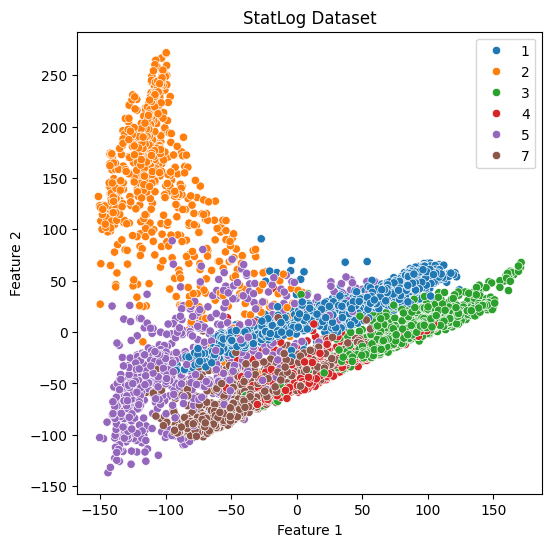}
    \caption{Statlog Dataset}
    \label{fig:statlog_dataset}
\end{figure}

Figure \ref{fig:statlog_dataset} shows the Statlog dataset after applying principal component analysis.


\end{document}